\documentclass[11pt]{article}
\usepackage{amsmath}
\usepackage{graphicx}
\usepackage{epsfig}
\begin{document}

\title{Some Advantages of a Local Realist, 3D Wave Soliton Approach to EPR}
\author{A. Laidlaw\thanks{e-mail: a\_laidlaw@today.com.au}}
\date{2001 October 28}
\maketitle

\begin{abstract}

Not only are the foundation theories mutually compatible, they are
also compatible with local realism once this concept is properly
formulated (without presuming atomism in addition to locality).
Relativity Theory is reconstructed in the context of a preferred
frame, but so as to secure the Relativity Principle in every
experiment except the measurement of the preferred frame, defined
by a null result for the dipole component of the Microwave
Background Radiation (MBR).  The 3D soliton approach to physical
modelling is found to be consistent with both Special Relativity
and Quantum Mechanical Nonlocality, exemplified by the EPR
paradox.

\end{abstract}

\section{Introduction}
\label{sec:intro}

Quantum Mechanics makes the fundamental assertion that we can
obtain the best possible predictions concerning the outcome of a
given physical situation by adopting certain mathematical
procedures (consistent with its axioms) in order to: a) encode
information known about the system into a "wavefunction"; b) chart
the evolution of this representation between preparation and
measurement; and c) extract predictions by re-expressing it as a
linear superposition of the eigenfunctions of simultaneously
measurable observables, and then applying the projection
postulate. Whilst there remain physical situations for which the
appropriate mathematical formalism remains the subject of debate
(eg continuous measurements; relativistic state reduction), and
whilst the future may prove different, there is a complete absence
of any evidence based challenge to this remarkable, information
theoretic approach.  All the evidence shows that we can succeed in
any practical goal by restricting ourselves to a consideration of
in-principle accessible information alone.

On the other hand, it so happens that quantum mechanical
predictions conflict with some deeply held metaphysical beliefs
concerning the nature of interaction, which are also frequently
associated with, but actually predate, the Special Theory.
Specifically, and quite independent of relativity, they conflict
with the commonly assumed paradigm of retarded interaction,
wherein interaction is mediated by the motion of some element of
the ontology ("Interaction Mediating Particles" - IMPs) between
primary, point particle participants.  The problem has now been
clarified, refined and subjected to experimental investigation by
virtue of several famous contributions especially those due to EPR
\cite{EPR}, which clearly exposed the basic conflict by showing
that quantum mechanics predicts influences between space-like
separated events, and Bell \cite{Bell}, which both made the matter
testable, and showed that it could not be resolved by recourse to
any Local Hidden Variables (LHV) method. Although there are a
couple of outstanding loopholes \cite{Clauser, Vaidman}, all the
experimental evidence \cite{Aspect, Scarani, Ou} firmly supports
the quantum mechanical predictions, leaving the physics community
with a new, and qualitatively different kind of challenge to all
those it has dealt with in the past.  The current debate is
dominated not by its quantitative aspects, which we may take to be
as given by Ordinary Quantum Mechanics (OQM), but rather by the
need of the community to understand how such facts could be
physically possible.  In short, the problem with EPR is unique in
that it lies not so much in the field of Physics, but rather in
that of Metaphysics.

An analysis by Percival \cite{Percival} has exposed an essential
feature of the EPR paradox \cite{EPR}.  Percival considers sets of
two (or more) EPR experiments in relative motion with respect to
each other, and so organised that the "inputs" (choices of the
measurement axis in a given arm) of one apparatus are determined
by the "outputs" (measurement outcomes along the chosen axis) of
the other.  Assuming only relativity of simultaneity and the
relevant quantum predictions,  he shows that it is possible, with
a judicious choice of the reference frame pertaining to each
experiment, to generate paradoxical temporal loops (such that an
input is equal to the opposite of itself) - a manifest and
in-principle testable contradiction that questions the notion of
"peaceful coexistence".

Since these two assumptions lead to a contradiction, at least one
of them must be false, and since the quantum predictions just
describe the experimental outcomes, we have to look closely at the
concept of relativity of simultaneity. Is it a completely general
"metaphysical truth" (as Physics presently has it), or merely an
artefact of observation, a matter of perspective? The introduction
of a preferred frame would go a long way to address the problem,
but this is an option seldom considered in any detail because a
preferred frame (or so it seems) must disrupt the mathematical
symmetries at the heart of all modern physics.

This article shows that there is at least one way to introduce a
suitable, experimentally observable, preferred frame so as to open
the door to EPR, whilst preserving these valuable symmetries and
retaining the 4-vector calculus. We cannot progress towards a
physical understanding of EPR without introducing some physical
content, so let us identify a single, radically simplifying,
ontological constraint to take the place of the usual
epistemic\footnote{i.e. both postulates relate to the facts of
observation} postulates underlying the Special Theory, and from
which these will be derived.

Radiation propagates at the characteristic velocity, c, whilst,
for matter, only speeds below the characteristic velocity can be
observed. The logical necessity of an energy equivalence between
these distinct classes of phenomena is already reflected in the
Theory, but this does raise a perplexing question about photon
absorption processes: How can the inherently propagative be
transformed into the inherently non-propagative?  A simple, but
highly productive, resolution of this is to deny the possibility,
and instead raise the matter-radiation energy "equivalence" to the
level of an identity by asserting that the energy constituting a
photon cannot be fundamentally different from that constituting a
massive body. In this case, the fact that the photon has a
characteristic velocity mandates that energy has a characteristic
velocity in all contexts, matter as well as radiation.  This will
be our central assumption: energy propagates at c.

We shall show that subluminal phenomena in a universe in which all
the "elements of reality" are dynamically constrained to constant
speed motion behave relativistically, whilst the converse, that a
universe that behaves relativistically has little choice but to be
constructed from elements constrained to constant speed, seems
 to be strongly implied.  After first identifying the Microwave
Background Radiation (MBR) as the only plausible candidate
preferred frame, and then discussing its essential properties, a
second motivation behind this assumption is found in Section 2 -
it binds the only suitable candidate preferred frame tightly to
Relativity Theory and therefore makes it highly relevant to the
ordinary practice of Physics.

As far as particle models are concerned, the dynamical constraint
places us in the domain of 3-dimensional, local realist wave
soliton models - particle models constructed from definitively
"physical" wave elements.  Several consequences of this general
approach are considered in Section 3, where the analysis is based
on the necessary conclusion that, when we consider physical,
momentum carrying entities (such as waves), then Conservation of
Momentum (CoM) can only mean that the momentum of a superposition
is given by a sum over its components' momenta. The usual form of
wave packet analysis (using a dispersion relation in the context
of infinitely extensive wave components constrained to propagate
in the same direction as the motion of the particle) will be found
to be non-physical. This direct approach to CoM in 3D soliton
wavepackets leads immediately to: a) the invariance of the
4-momentum; b) an internal "clock" concept and the value of its
time dilation parameter; and c) the elliptical transformation of
the fields of a moving charged particle (a surrogate for the
Lorentz-Fitzgerald contraction).

The result is that we have the usual relativistic symmetries,
embedded in a preferred frame.  The final task (Section 4) is to
reconsider our understanding of the EPR paradox from this new
perspective. We shall find a significant loophole in EPR's
sufficiency condition for an "element of reality".  In the
physical wave soliton context, where the localised image of a
"point-particle" results from a Fourier superposition of
inherently distributed components, it is wrong to assume that
elements of reality must be co-located with the observations to
which they correspond. From this it is argued that EPR does not
conflict with local realism (the combination of proximate
causation with a "no superluminal movements" contraint), only with
point or point-like local realist models (i.e. LHV models), and
that the inherently extended nature of wave phenomena is logically
sufficient to explain all the known experimental evidence without
invoking instantaneous action at a distance or indeed any
phenomenon that moves faster than light.\\

\section{The Preferred Frame}

\subsection{Specification}

If we are to include this controversial idea amongst the
foundations upon which physical theory is built, then the
preferred frame must be unequivocally experimentally observable.
In addition, any solution relying on observations over large
(galactical) distances is fraught with difficulties from the
outset since we cannot assume the homogeneity over such great
distances of any underlying reference system.  So, let us limit
the range of acceptable proposals to those that can be determined
exclusively by local measurements. Only one preferred frame has
been proposed that conforms to this criterion. It is the frame of
reference defined by a null result of a measurement of the dipole
component of the Microwave
Background Radiation (MBR) \cite{Smoot, Peebles3}.\\

\subsection{Key Properties of the Preferred Frame}

Consider an observer remote from all massive objects and equipped
with a suitable directional detector for the MBR temperature. He
first finds that the temperature distribution is in general
anisotropic, but that it can be expressed as the sum of an almost
perfectly isotropic distribution and a dipole component (which is
of order one part per thousand for the earth, corresponding to a
speed of approximately 350 Km/Sec) \cite{Longair, Weinberg}.
Varying his own condition of motion in a controlled manner and
repeating the experiment several times, he finds that the strength
and direction of the dipole component depends upon his condition
of motion, and that there exists a particular condition of motion
for which there is no dipole component.  He can achieve this
condition of motion by accelerating towards the minimum of the MBR
temperature distribution in  3-space.

Since, prima facie, the MBR radiation is expected to be isotropic,
he infers that the dipole component of the MBR temperature is just
the doppler shift induced by his own motion.  When no dipole
component can be observed, the doppler shift from his own motion
is equal in all directions, and therefore equal to zero. He
recognises that the same must be true, not only for MBR photons,
but for all photons in his local space, emitted as well as
absorbed. Amongst all observers, his measurements of all photon
momenta in general are privileged in not being influenced by his
own motion, so anything that propagates at c is referenced to the
MBR preferred frame\footnote{including EPR experiments with
photons.}.

In order to extend the applicability of the MBR preferred frame to
cover massive bodies in addition to the radiation, he reasons that
this can follow if, and only if, his measurements of all momenta
(massive particles as well as radiation) are similarly privileged,
which   in turn follows immediately if the massive particles are
formed as superpositions of generalised photonic waves. Since the
components of such superpositions are referenced to the preferred
frame, the superposition as a whole is automatically similarly
referenced provided only that momentum is conserved. The
mathematical forms of such waves are irrelevant because logic
requires only that momentum propagates at the characteristic
velocity and is conserved.

\subsection{Implications for Modelling Massive Particles}

Massive bodies are usually well-localised, whereas propagative
phenomena are usually well-distributed, so we invoke the Fourier
principle to explain the localised appearance as an interference
phenomenon amongst multiple, distinct, extensive, interpenetrating
waves.

Photonic waves all propagate at the same speed, whilst the
material body has variable speed.  Therefore the direction of
propagation of a given wave component must vary in time and so
cannot be fixed in the direction of motion of the particle (as is
common in the usual form of wave packet analysis). For example, a
rest particle might be represented by two physical wave components
of equal momentum which always propagate in opposite
directions\footnotemark  (or in a variety of other ways).
\footnotetext{which is to say they are spinning about each other.}

Material bodies persist in a self-similar condition throughout
extended periods of time, whereas a superposition of real,
physical, photonic waves would immediately dissipate.  It follows
from this that there must be interaction between the wave
components such that they execute bounded motion and so remain
associated as a group\footnotemark. This seems to be, but is in
fact not, a new assumption. If we contemplate only photonic
elements, and recall that we observe interactions amongst the
massive particles, then interaction amongst the constituent waves
of different particles is absolutely implied, and so, therefore,
is interaction between the different constituent waves of the same
particle.

\footnotetext{Typically introduced into soliton analyses by virtue
of a non-linearity in the medium.}

This basic concept of distributed wave-wave (or field-field)
interaction is a complete departure from the usual metaphysical
framework where the problem is stated in terms of separated point
particles which can affect each other across a large distance only
by the exchange of (retarded) IMPs.  In the distributed case, no
point-to-point relationships need to be calculated because every
object is present at every point, at least in principle.  The
interaction occurring at some place and time depends upon those
parts of the various objects (e.g. their respective field
variables) that are co-located at that place and time, and the
total interaction is an integral over all space of the local
interactions.  There is no need for action at a distance, retarded
or instantaneous.

\section{Lorentz Invariance of Photonic Systems}
\label {sec:Lorentz}

Given that the several, standard relativistic wave equations
(including especially the Helmholz and Dirac equations) all
feature the same characteristic velocity, the suggestion that
systems of photonic waves exhibit Lorentz Invariance is no great
surprise.  In this section, we derive the usual relativistic
invariances directly, without invoking any wave equation which
would limit the scope of the results to the particular equation(s)
at hand, by showing that any field of 3-momentum subject to CoM
and the dynamical constraint obeys the Special Theory. Wave
momentum behaves differently from the linear momentum concept
assumed by Galileo and Newton - it behaves like relativistic
momentum.

The critical assumption upon which this relies is that, having
once asserted that the superposed wave objects are physically
real, we must allow that the wave momentum carried by each
component is equally real, so the particle momentum must be a
(vector) sum over components' wave momenta. This requires some
explanation because it is not usually valid in a wave packet
analysis, but let us first emphasise that this has nothing to do
with quantum mechanics, where the components are not even
considered to be physical, and where CoM between wave components
and the particle turns on the probability weights Ordinary Quantum
Mechanics (OQM) attaches to each of its wave components in the
momentum basis. OQM provides good quality predictions, period.
Although wave packet analyses originated in physical situations,
the wave components introduced by Fourier analysis are typically
constrained to propagate in the same direction as the particle.
This produces a 1-dimensional rather than a 3-dimensional image,
dispersion is mandatory, and the wave components must be thought
of as infinitely  extended in the direction of motion, so that the
representation of the original, finite object is itself infinite
and therefore non-physical. Since the components continuously
slide past the localised image, CoM between such abstract
components and the superposition must be explicated using an
appropriate wave equation.

By contrast with this, with bounded motion under the dynamical
constraint introduced above, the momentum that forms a given wave
component constantly changes direction (due to interaction with
the other wave components), so that its time average velocity is
equal to the particle's group velocity, under which conditions CoM
should be applied directly.

\subsection{Invariance of the 4-Momentum}
\label{ss:mechanics}

Consider the following general, multi-component form for a  wave
packet:

\begin {equation} \label{eq:packet}
\Psi(\underline{\textbf{r}},t))=\begin{array}{c}
  \lceil \psi_{1}(\underline{\textbf{r}},t) \rceil \\
  :\\
  \psi_{i}(\underline{\textbf{r}},t) \\
 : \\
  \lfloor\psi_{N}(\underline{\textbf{r}},t)\rfloor \\
\end{array}
\end {equation}

Where each $\psi_{i}$ describes a momentum carrying wave component
in 3-space + time (and so includes its direction of propagation,
the unit vector $\widehat{\theta}_{i}$). Throughout this article
we shall use lowercase symbols to refer to wave components, and
uppercase to refer to superpositions as a whole. Let the momentum
carried by the $\textrm{i}^{th}$ component be
$\underline{\textbf{p}}_{i}$. Suppressing the functional forms of
the various components, we can, by our CoM assumption, always
write for the particle momentum:

\begin {equation} \label{eq:P}
\textbf{\underline{P}}=\sum_{i=1}^{N}\textbf{\underline{p}}_{i} =
\sum \textbf{\underline{p}}_{i}=\sum
\textrm{p}_{i}\widehat{\theta}_{i}
\end {equation}

\noindent Where: \\
$\textbf{\underline{P}}$ = Group (or particle) momentum\\
$\textbf{\underline{p}}_{i}$ = momentum carried by the
$\textrm{i}^{th}$ component\\
$\textrm{p}_{i}=\|\textbf{\underline{p}}_{i}\|$\\
N = number of components in the particle representation. Since the
summation range is always from 1 to N, it will be omitted from
here on.

Now let us also put c=1, again by choice of units, so that the
energy, $\textrm{e}_{i}=\hbar\omega_{i}$, of the $\textrm{i}^{th}$
component is equal to the scalar value of the momentum carried,
$\textrm{e}_{i}=c\|\textbf{\underline{p}}_{i}\|=\textrm{p}_{i}$,
and so the total energy of the superposition is given by:

\begin {equation} \label{eq:E}
\textrm{E}=\sum\textrm{e}_{i}=\sum\textrm{p}_{i}
\end {equation}

Noting that component velocities, ($\textbf{\underline{v}}_{i} =
\textbf{\underline{p}}_{i}/\textrm{p}_{i}$), all have unit modulus
but variable orientations, the group velocity (consistent with our
CoM assumption) is a momentum weighted average:

\begin {equation} \label{eq:V}
\textbf{\underline{V}}=\frac{\sum\textrm{p}_{i}\textbf{\underline{v}}_{i}}{\sum\textrm{p}_{i}}
\end {equation}

\noindent Where \textbf{\underline{V}} = Group velocity. The
modulus of group velocity, V, is a real number in the range [0, 1]
(this is $\beta$ in most texts).  Define the effective mass from
the usual relation between momentum and velocity:
\begin {equation}     \label{eq:P=meV}
\textbf{\underline{P}} = \textrm{m}_{e}\textbf{\underline{V}}
\end {equation}
So, from equations \ref{eq:E}, \ref{eq:V} and \ref{eq:P=meV}:
\begin {equation}   \label{eq:me}
\textrm{m}_{e}=\frac{1}{c}
\sum\|\textbf{\underline{p}}_{i}\|=\sum\textrm{p}_{i}=\textrm{E}
\end {equation}

\noindent The rest mass, $\textrm{m}_{0}$, is just the effective
mass at zero group velocity.  The above definitions will turn out
to be good for all observers, but to make it clear that we are in
no way assuming the result, let us begin by restricting the
analysis to an observer at rest in the MBR frame. As the momentum,
\underline{\textbf{P}}, of a typical superposition varies in
response to an interaction, the question arises how changes in the
group momentum become distributed amongst the components. As far
as these individual $\dot{\textbf{\underline{p}}}_{i}$ are
concerned the following conditions are required if the superposition
of momenta is to be linear as CoM mandates: \\

a)   $\dot{\textbf{\underline{p}}}_{i}\propto \textrm{p}_{i}$ (In
order to preserve superposition of
components)    \\

b)  $\dot{\textbf{\underline{p}}}_{i}\propto
\dot{\textbf{\underline{P}}}$ (In order to preserve superposition
of interactions) \footnote{This condition can (and probably
should) be relaxed without affecting the analysis by replacing the
LHS with its time average or an expectation.  The main point is
that, by definition, any  components of the
$\dot{\textbf{\underline{p}}}_{i}$ that are transverse to
$\dot{\textbf{\underline{P}}}$ sum to zero, so we need deal only
with the parallel components.}\\

c)
$\dot{\textbf{\underline{p}}}_{i}\propto\frac{1}{\textrm{m}_{e}}$.
Interactions are usually calculated on the basis of an invariant
"interaction" property (cf: "electric charge" is the interaction
property of the Electromagnetic interaction.) Since the sum of
scalar momenta (the energy of the superposition) varies under
interaction, we also require that
$\dot{\textbf{\underline{p}}}_{i}\propto\frac{1}{\textrm{m}_{e}}$
in order to retain the invariance of the interaction property.\\

\indent {d)  $\dot{\textbf{\underline{p}}}_{i}\neq
f(\frac{\textbf{\underline{p}}_{i}}{\textrm{p}_{i}})$
($\frac{\textbf{\underline{p}}_{i}}{\textrm{p}_{i}}$ is a unit
vector, $\widehat{\theta}_{i}$ in the direction of propagation.)
Since the various $\dot{\textbf{\underline{p}}}_{i}$ are parallel,
all the components of a superposition rotate towards the same
direction under interaction. Therefore
$\dot{\textbf{\underline{p}}}_{i}$ must be independent of
$\widehat{\theta}_{i}$ to preserve the interaction property.\\

\noindent Summarising, the distribution of changes to the group
momentum amongst components is governed by:

\begin {equation}    \label{eq:dp}
\dot{\textbf{\underline{p}}}_{i}\propto
\dot{\textbf{\underline{P}}}(\frac
{\textrm{p}_{i}}{\textrm{m}_{e}})\Rightarrow\dot{\textbf{\underline{p}}}_{i}=
\dot{\textbf{\underline{P}}}(\frac
{\textrm{p}_{i}}{\textrm{m}_{e}})
\end {equation} \\
The effect of Equation \ref{eq:dp} is just to apportion the
interaction property amongst the components of the group whilst
conserving it for the group as a whole.    In principle, one
cannot rule out other kinds of dependency (especially upon the
phases of components).  However the equality must apply to an
average over a time interval sufficient to measure the group
momentum and/or its rate of change.

In contrast to the classical idea, where a force causes a change
in the condition of motion of an otherwise unchanged particle, an
increase in the momentum carried by a wave propagating at constant
velocity implies a substantive change to the wave itself (a change
in amplitude or frequency, for example)\footnote{It will become
clear later that the change is, in this case, a change in
frequency}, as opposed to a mere change in its condition of
motion.  Now, consider a superposition for which
$\textbf{\underline{P}}=0$ at $\textrm{t}=0$, in the MBR frame.
From Equations \ref{eq:P}, \ref{eq:me}, and \ref{eq:dp}
above, we have: \\

\noindent $ \textbf{\underline{P}}=\sum \textbf{\underline{p}}_{i}
=\int\dot{\textbf{\underline{P}}}\textrm{dt}= \int \sum
\dot{\textbf{\underline{p}}_{i}}\textrm{dt}$ \hspace{.5cm};
\hspace{.5cm}
$\dot{\textbf{\underline{p}}}_{i}=\dot{\textbf{\underline{P}}}(\frac{\textrm{p}_{i}}{\textrm{m}_{e}}
)$  \hspace{.5cm}  and \hspace{.5cm}$\textrm{m}_{e} =\sum
\textrm{p}_{i} \Rightarrow \dot{\textrm{m}_{e}} = \sum
\dot{\textrm{p}}_{i}$
  \\

\noindent Where $ \dot{\textrm{p}}_{i}$ is the component of
$\dot{\textbf{\underline{p}}}_{i}$ parallel to
$\textbf{\underline{p}}_{i} $. One readily finds that
$\textrm{m}_{e}\dot{\textrm{m}_{e}}=\textbf{\underline{P}}\cdot\dot{\textbf{\underline{P}}}
$.  Integrating, we get:

\begin{equation}   \label{M2}
\textrm{m}^{2}_{e}= \textrm{P}^{2} + \textrm{m}^{2}_{0}
\end{equation}
Which is equivalent to stating that the norm of the
energy-momentum 4-vector of a particle is invariant for observers
in the MBR.  Substituting equation \ref{eq:P=meV} in this gives:

\begin{equation}   \label{P=gammam0V}
\textbf{\underline{P}}=\frac{\textrm{m}_{0}\textbf{\underline{V}}}{\sqrt
{1-\textrm{V}^{2}}}=\gamma \textrm{m}_{0}\textbf{\underline{V}}
\end{equation}

\noindent   Since c=1 by choice of units.  Although this
determines mechanics within the selected frame, momentum involves
both length and time, so reconstructing the Lorentz Transformation
between frames requires that we now identify at least one of these
by itself. The following subsections develop both time and length
transformations separately, however the analysis of length
contraction (Subsection \ref{ss:transform}) necessarily touches on
Electromagnetics, so we take time dilation first.

\subsection{Time dilation} \label{ss:time}

As is evident from Equation \ref{eq:V}, spatial correlation
amongst the directions of propagation, $\widehat{\theta_{i}}$, of
the components of a superposition is necessary if it is to have a
non-zero group velocity (again, let us establish the result in the
MBR frame first). Higher degrees of correlation correspond to
greater relative velocities. As the group velocity approaches
the speed of light, the components approach the parallel: \\

As $  \textrm{V}\rightarrow 1,\;\; \widehat{\theta_{i}}\rightarrow
\widehat{\textbf{V}}$ For all i.\\

But, if all the components of a group were exactly parallel to
each other, no changes would arise in the spatial configuration of
the group as it moved through our observer's frame of reference.
Considering such a situation he must conclude that nothing ever
happens in the inertial frame of the group.  So, just as
correlations amongst the $\widehat{\theta_{i}}$ are necessary for
movement of the group, decorrelations are necessary for it to
evolve internally.  As the velocity of a group increases, its rate
of internal evolution reduces. Changes in the spatial
configuration of components with respect to each other (internal
evolution) are thus associable with the passage of time, so the
internal movements form a velocity dependent clock.

Each component contributes to the evolution of the group spatial
configuration by virtue of its motion relative to the group (its
"internal" motion).  So, if we write, for the $\textrm{i}^{th}$
component:

\noindent $\textbf{\underline{v}}_{i} = \textrm{c}
\widehat{\theta_{i}} (=
\textbf{\underline{p}}_{i}/\textrm{p}_{i})$ then the corresponding
internal motion is given by:

\begin{equation}    \label{eq:vzi}
 \textbf{\underline{v}}_{zi}=c \widehat{\theta_{i}} - \textbf{\underline{V}}
 =\widehat{\theta_{i}}- \textbf{\underline{V}}
\end{equation} \\
There is a meaningful comparison between this simple expression
and the motion of the electron as described by the Dirac Equation
\cite{Dirac}, for which the velocity operator,
$\overrightarrow{\alpha}$, has constant modulus, c \cite{Breit}.
Although the instantaneous speed of the electron is
 constant in the theory, this is usually thought of in two
 parts, the group velocity $\underline{\textbf{P}}/\textrm{H}$, and
 a high frequency ($\sim 2H/h$), small amplitude ($\sim \hbar/2\textrm{mc}$),
  internal oscillatory motion, commonly known as the
zitterbewegung \cite{schroedinger, Dirac3}.  Equation
 \ref{eq:vzi} describes the internal motion of a component of a generalised
 photonic superposition, and in this sense corresponds to the zitterbewegung.
Both the zitterbewegung and the $\textbf{\underline{v}}_{zi}$
 scale with the group velocity according to the usual time
 dilation parameter, $\gamma$.

With respect to the zitterbewegung, the result follows from the
equation for the time dependence of the velocity operator in the
Heisenberg representation of the Dirac theory \cite{Messiah}:

\begin{equation}    \label{eq:alpha}
\overrightarrow{\alpha}(\textrm{t})=
(\overrightarrow{\alpha}(\textrm{0})-\frac{\textbf{p}}{\textrm{H}})\exp{(-2\textrm{iHt})}+
\frac{\textbf{p}}{\textrm{H}}
\end{equation} \\

In which $\textbf{p}$ and $\textrm{H}$ are both constants, so $
\textbf{p}/\textrm{H} = \textbf{V}_{\textrm{g}} =
\textrm{constant}$. The quantum mechanical expectation of the
zitterbewegung, the first term on the RHS, is then
$\frac{<\Psi\mid(\overrightarrow{\alpha}(\textrm{0})-\textbf{V}_{\textrm{g}})\mid\Psi>}
{<\Psi\mid\Psi>}$ which varies with the group velocity,
$V_{\textrm{g}}$, as $\sqrt{1-V^{2}_{\textrm{g}}}$, since
$\overrightarrow{\alpha}$ has real eigenvalues.

To establish the same result for the
$\textbf{\underline{v}}_{zi}$, note that the term "internal
motion" has no meaning except in the context of a superposition.
We must ensure that two representations which are identical in all
respects except that, in the direction $\widehat{\theta_{i}}$, one
has, say, $\textrm{p}_{i}$ components of strength unity, whilst
the other has a single component of strength $\textrm{p}_{i}$, are
treated equivalently. So the necessary measure to connect these
 zero mean components of the internal movement with the evolution of the
pattern formed by the whole superposition is a momentum weighted
standard deviation.  Let us define:

\begin{equation}  \label{eq:VZ}
 \textrm{V}_{Z}=\sqrt{\frac{\sum \textrm{p}_{i}
 \textrm{v}_{zi}^{2}}{\textrm{m}_{e} }}
\end{equation} \\

\noindent It is readily shown (expand the square as the dot
product of $(\widehat{\theta_{i}}-\textbf{\underline{V}})$ with
itself) that:

\begin{equation}
\textrm{V}_{Z}=\sqrt{1-\textrm{V}^{2}}
\end{equation}
is the time dilation factor for superpositions of propagating
waves.

The argument from internal atomic processes to real world clocks
has long been established \cite{Bohm}, and has been the subject of
exhaustive empirical review \cite{Hafele, Kundig, Ives}, so we
find:
\begin{equation} \label{dtdtau}
 \frac{\Delta t'}{\Delta t}=\frac{1}{\gamma}
=\textrm{V}_{Z}
\end{equation}\\
\noindent Where  $\Delta t'$ is the interval observed to pass on a
moving clock corresponding to $\Delta t$, the interval observed to
pass on a stationary clock.

This reduction in the rate of internal evolution, or time
dilation, is a direct consequence of the constraint to constant
velocity propagation:-  To whatever extent the motion of a given
wave component contributes towards transporting the superposition
through space, it is unavailable to contribute towards its
(temporal) evolution in situ and vice versa.

\subsection{The Electromagnetic Fields of a Moving Particle}
\label{ss:transform}

The preceding sections have shown how analysis of spatial
correlations amongst the components of a photonic superposition
leads to characteristic relativistic behaviour.  Here, we consider
how the correlations develop as a superposition is Lorentz
boosted. The result is a relation connecting the group velocity of
a superposition with the shape of its momentum flux distribution
in 3-space.

This relation echoes the transform that connects the
Electromagnetic fields of a moving particle to those of a
stationary particle.  Of course it is well known that this
transform describes a compression of the Electromagnetic fields of
a charged particle in the direction of its motion, which equates
to a Lorentz contraction of dimensions in the same direction
\cite{Bohm}.

\subsubsection{Numerical Analysis}

We take the case of a rest particle with an isotropic distribution
of the momentum flux, and consider how it appears to other Lorentz
observers.  This distribution is defined by the condition that the
expected value of the momentum flux density in one direction is
equal to that in any other direction (which obviously gives us
$\underline{\textbf{P}}=0$).  In order to illustrate the
(elliptical) distortion of this distribution when considered from
various reference frames, the equations of Subsection
\ref{ss:mechanics} have been analysed numerically. The result,
shown in figure \ref{fig1}, is a series of ellipsoids of
revolution whose long axes lie in the direction of motion, and
whose eccentricity increases with the group
velocity.\\

\subsubsection{Calculation of the Momentum Flux Distribution}

This can be analysed directly if we replace the spherically
symmetric distribution by a superposition of balanced pairs of
waves (waves of equal but opposite momentum). If we put N = 2 in
the analysis of Sub-Section \ref{ss:mechanics}, it can be seen
that a balanced pair of waves contributes to the rest mass, but
not to the particle momentum of a larger superposition. However it
may be oriented, such a pair transforms (under interaction or,
equivalently, a change of referential) such that its contribution
to the total energy of a moving superposition is always in
proportion to its contribution to the rest mass energy, so a
balanced pair transforms independently of the rest of a
superposition.

Consider such a pair of waves arbitrarily oriented with respect to
the velocity separating two inertial frames. Figure \ref{fig2}, in
which the X-axis is selected to lie along the direction of the
velocity separation, depicts the situation. Fig. \ref{fig2}(a)
shows a pair of waves, ($a_{0}, b_{0}$) the sum of whose momenta
is zero, whilst fig. \ref{fig2}(b) shows the same pair from an
inertial frame in which the group velocity of the superposition is
$ \textbf{\underline{V}}=\textrm{V}_{x}\widehat{\textbf{i}}$.

Since we are concerned only with the ratios, let $
\textrm{a}_{0}= \textrm{b}_{0} = 1 $ (in momentum units). Then:\\

 \noindent $ \textrm{a} + \textrm{b} = 2\gamma$ ;
  $ \textbf{\underline{a}} + \textbf{\underline{b}}
  = 2\gamma\textrm{V}_{x} \widehat{\textbf{i}}$ \\
 $\textrm{a}^{2} = \textrm{r}^{2} + (\gamma\textrm{V})^{2} +
2\gamma\textrm{Vr}\cos\Theta$ \\
 $ \textrm{b}^{2} = \textrm{r}^{2}
+ (\gamma\textrm{V})^{2} - 2\gamma\textrm{V}\textrm{r}\cos\Theta$
\\

\noindent Upon eliminating a and b from this, it is found that:
\begin{equation}   \label{eq:transform}
  \textrm{r}  = \frac {1}{\sqrt{1-\textrm{V}^{2}\cos^{2}\Theta}}
\end{equation}  \\

This transform connects the expected value of the momentum
instantaneously propagating in a given direction with the group
velocity for superpositions of generalised photonic wave
components.

\subsubsection{Compression of the $1/r^{2}$ fields}

Now, let us bring the Electromagnetic field into the discussion by
positing that attributable to each photonic wave component is an
associated Electromagnetic field\footnotemark, and that it is the
superposition of these that appears to us as the fields of the
particle, described by the interaction property, "charge", Q.
  Allocating this property to wave components in accordance with
equation \ref{eq:dp} gives:

 \footnotetext {i.e. the EM fields are a property of the waves, to which
 we have granted prior status.  We do not suggest that the EM field is the
 total reality.}

\noindent $ \textrm{Q(V)}=\textrm{Q(0)}\Rightarrow\textrm{q}_{i}=
\textrm{Q}\textrm{p}_{i}/ \textrm{m}_{e}=
\textrm{Q}\textrm{p}_{i}/ \gamma\textrm{m}_{0}.$  Following this
through introduces an extra factor of $\gamma$ in the denominator
of the RHS of equation \ref{eq:transform}, upon which it becomes
the result usually calculated to transform the Electromagnetic
fields of a rest particle into those of a moving particle
\cite{Konopinski2}, except that $\cos\Theta$ has replaced
$\sin\Theta$.  Poynting's Theorem \cite{Poynting} has it that the
Energy flux density vector \textbf{S}, is parallel to the momentum
carried by the Electromagnetic field, and in the direction of the
cross-product \textbf{E x H}, so \textbf{E} and \textbf{H} are
both transverse to \textbf{p}, and we should expect $\cos\Theta$
to replace $\sin\Theta$ in a transform expressed in terms of wave
momenta.

Finally, forming this connection to Electromagnetics has ensured
that the physical wave elements, whilst unbounded, are indeed
finite.

\subsection{The Relativity Principle}
\label {ss:Special}

Given equations \ref{P=gammam0V} and \ref{dtdtau}, the observer in
the MBR can use standard methods to reconstruct the perspective
encountered by any other inertial observer.  In doing so, he will
deduce the Lorentz Transform, the Relativity Principle and the
constant observed velocity of radiation without any heuristic
redefinition of momentum, and without needing the Ives-Stillwell
experiment \cite{Ives} to confirm the separation of time dilation
from length contraction.  The Relativity Principle states that all
experiments work the same for all observers. Since the
experimental fact is that different observers do not get the same
result when measuring the MBR temperature distribution, the wave
soliton approach includes new facts which cannot be accounted for
within the usual interpretation. Doesn't this exception undermine
the Relativity Principle?

On the contrary, measurements on the MBR are the exception that
proves the rule: Photonic motion both defines the preferred frame
and forms the basis for inherently relativistic particle models.
Consequently there is no reason to reject the Relativity Principle
in any other context.  As far as the empirical Physics goes,
Einstein's original argument, although it was formulated within a
particle theoretic framework, did not depend upon it. Now, it is
widely presumed that this argument precludes the preferred frame
concept, but this is simply not the case - the fact that the
Special Theory does not \textit{require} one neither proves nor
disproves the existence of a preferred frame, so Relativity Theory
is silent \cite{BohmHiley}.  Philosophy (Machian positivism), not
Physics, provided what has long been the decisive argument - that
all the elements of a theory ought to be observable, at least in
principle. Although this remains sound, the positivist error lay
in assuming that what was then unimaginable would remain forever
impossible - a century ago it was impossible to predict the
discovery of the MBR. Since we do have an observable preferred
frame, the philosophical argument that excluded the preferred
frame concept on the basis of its vaunted unobservability ought
now to be inverted - we require a theory that includes all the
observables. The present interpretation categorically fails to
achieve this goal.

The Relativity Principle was originally postulated on essentially
aesthetic grounds. Now, it has been deduced from one of the surest
observations in all of Physics - there are phenomena that
propagate at the characteristic velocity. Rather than introducing
new concepts, we removed the concept of inherently sub-luminal
motion, then re-synthesised it from superposed photonic movements.
Perhaps to deny the existence of matter qua material substance is
a drastic step, but modern physics has taught us to think of
material particles as dynamic systems, and has eradicated the
notion of substance as well as any concept of persistence of
identity \cite{Teller}, so what real evidence is there for a
non-propagative energy? Where is the conceptual economy in a
Physics with two distinct forms of the same thing? How should
something which moves by its nature be transformed into something
not inherently motional?  The only support for the complications
introduced by the idea of matter as distinct from radiation lies
in the notoriously unreliable "common experience".

Before returning to the EPR paradox, it is worth noting two recent
papers by H.Y. Cui \cite{Cui2, Cui1}, in which a series of key
results, from the structure of Electromagnetics to the
Klein-Gordon and Dirac equations, as well as the Schwarzchild
metric tensor, have been deduced from the constancy of the
4-velocity.  We have shown that the invariance of the 4-velocity
is implied by the invariance of the 3-velocity, and from this it
follows that this significant body of knowledge can be inferred
from a single observation.

\section{Understanding EPR}
\label{sec:EPR}

Percival's temporal loop paradox is resolved - relativity of
simultaneity is in the eye of the beholder, and there is a
physically meaningful sense in which quantum mechanics relies upon
the distinction between what is immediately presented to observers
and what actually is.  "Wave function collapse" is in the MBR
frame.  Although constructing Relativity Theory around a preferred
frame opens the door to EPR, there remains, amongst the many
different lines of argument in the discussion of EPR, two
available paths in logic capable of explaining these experiments
physically.

These are superluminality and nonseparability arguments, as
epitomised by those of Redhead \cite{Redhead}. Whilst
superluminality seems readily intelligible, the somewhat subtler
nonseparability arguments are all too commonly dismissed as
obscure, philosophical, even non-physical.  It is the intention in
the balance of this article to clarify the reasoning involved by
placing it in a specific physical context, namely local realist
wave theories. We shall expand upon Redhead's central
philosophical conclusion, note some experimental instances, and
finally describe a typical EPR experiment from the wave
perspective.

First however, let us assess the relative merits of these two
proposals.

\subsection{Superluminality or a Local Realist Wave Theory?}
\label{ss:compare}

Within the context of the wave mechanical interpretation of
Relativity Theory provided above, the particles of matter are
thought of as 3-dimensional wave solitons.  Superluminality is
then the proposition that, in addition to this primary wave
ontology, there is a second ontological form (such as the
superluminal shock wave, eg \cite{SL1}) that mediates long range
interactions.

Refining the timing windows in EPR experiments can never eliminate
the possibility of superluminal interaction mediation, so this
proposition isn't falsifiable. On the other hand, it would
undermine the nonseparability case if such a refinement ever led
to a failure of the quantum predictions.  Physics (at least to
date) has always been essentially epistemological, and quantum
mechanics' central assertion is that this is good enough for all
practical purposes. Superluminality substantially undermines both
this assertion and the structure of Relativity Theory. To the
extent that there is an alternative, should one seriously
entertain an unobserved, brand new class of phenomenon that denies
the validity of an extensively tested, theoretical centrepiece
combining elegance with unprecedented predictive power, to which
there is no hint of any practical exception, and whose only
downside is the fact that it is difficult to understand why it is
so?

It will of course be observed that this article, based on an
assumption about physical reality, crosses the boundary at least
into the metaphysical sub-category of ontology.  However, human
beings (physicists included) DO seek to understand the world, and
will continue to demand that it have rationally comprehensible
mechanisms.  The information theoretic approach to physics can
never, in principle, assist us in that area.  With EPR, we already
have a working theory, but just can't decide whether or not it is
philosophically acceptable. Apart from the deduction that wave
function collapse can be evaluated in the MBR frame, this article
makes no new physics as such, but its central purpose is not to
replace either of the foundation theories, merely to point out
that they are compatible with at least one self-consistent,
realist framework. Epistemological physics has advanced to the
point where it actually sheds light on genuine metaphysical
problems, and there can be useful feedback, but these distinct
philosophical categories should still remain separate.

 Finally, superluminality retains the
retarded interaction paradigm which, though initially attractive,
raises too many problems. When it has been faithfully implemented
it becomes unintelligible and analytically disastrous (as in the
case of the pre-acceleration induced by the Abraham-Lorentz
self-force \cite{Konopinski}) - a central problem being: When the
(virtual) IMP (or worse still, field, as in Classical
Electrodynamics) has transferred its momentum to the target, how
and when is (or was?) the reaction communicated back to the
source?  Anyone who doubts the severity of this problem should
study the history of the failure to solve the two
body problem in Classical Electrodynamics.\\

\subsection{Philosophical Review of the Nonseparability Argument}
\label{ss:review}

Redhead's central conclusion is that Ontological Locality (OLOC)
does not of necessity imply Epistemological Locality (ELOC). An
alternate statement of this is that local realism (a philosophical
construct which maintains that no part of the ontology moves
faster than light) does not imply the Principle of Locality (which
erroneously insists that events at one place cannot influence
those at another, remote location within the light time).

We can begin to make sense of this initially bizarre conclusion by
considering the specific experimental situation discussed in the
EPR paper \cite{EPR}, in which a particle pair, having become
entangled in position-momentum, subsequently became well
separated. Measuring the position of one member of the pair
enables quantum mechanically a prediction with certainty
concerning the position of the other, which is to say the result
that would be obtained should the second particle's position be
observed. Noting that the "position of the particle" (an
observation) is not the same as "the particle" (a "thing in
itself") we are entitled to question the inadvertent and almost
universal assumption that the elements of reality that
"correspond" to an observation are necessarily co-located with it.
A trivial example demonstrates the alternative.

In a typical small tornado the most readily observed phenomenon is
a well-localised tower of dust in the distance.  But it is well
understood that this is caused by the distributed system of winds
in the wider vicinity of the tornado.  The winds suck the dust
into the air.  The distributed gives rise to the local.  To
suggest that the dust causes the wind would be absurd, but this is
the position we have adopted with respect to sub-atomic particles.

Once we combine this error in logic with the condition of
proximate causation (which remains vital), all long range
interactions must be retarded relative to the "body" of the
particle. On the other hand, if we recognise that localisation is
more reasonably seen as an effect caused by distributed field
"elements", the whole question of retardation becomes moot.  In
the context of wave soliton models, the retarded "attached field"
\cite{Konopinski} assumption makes no sense at all.  The essence
of the soliton approach is that interference amongst distributed
components gives rise to a well localised "image", so the more
reasonable assumption would seem to be an "attached particle".

This error of presuming the causal relationship between a
particle's "body" and its distant "fields" is closely related to
another difficulty common in the Physics literature, namely the
tendency to equate "realism" with "atomism".  Again, an example
(from the recent literature) illustrates the point:

\begin{quote}

".... the realist philosophy, which claims that the reason why we
see macroscopic objects as having definite forms and definite
localisations in space is that they really exist as such, quite
independently of us, that is, of our sensorial and intellectual
equipment.  We see them at definite places because they are at
definite places."  - B. d'Espagnat \cite{d'Espagnat}

\end{quote}

Now, whilst this is one realist formulation, it is far from the
only possible formulation of the fundamental realist proposition.
We are led to formulate an alternative that is more compatible
with a wave theoretical approach to realism, viz:

\begin{quote}

The reason why we see macroscopic objects as having definite forms
and definite localisations in space is that there exists, quite
independently of our sensorial and intellectual equipment, a
unique, objective physical reality underlying, and causative of,
our sense experience.

\end{quote}

This formulation, whilst still realist (as opposed to idealist,
instrumentalist, positivist and so on) conforms with d'Espagnat's
later statement that:

\begin{quote}

"most contemporary philosophers take it as more reasonable to
consider that "the fact that we perceive such "things" as
macroscopic objects lying at distinct places is due, partly at
least, to the structure of our sensory and intellectual
equipment."" \cite{d'Espagnat}

\end{quote}

 And so:

\begin{quote}

 "We should not therefore take it as being part
of the body of sure knowledge that we have to take into account
for defining a quantum state." \cite{d'Espagnat}

\end{quote}

 With which, noting that the word
"it" in this quote includes the association between the location
property and the "thing in itself", we have no issue.  Indeed, it
is central to the nonseparability argument that we reject such an
association, and instead assume that the  reality underlying the
observation of an (unextended) location property is itself
inherently extended.  This is the converse of atomism, whose
central idea is that the reality actually subsists where we
identify it to be, a position which contradicts either locality or
quantum mechanics or both. Since the observation of point-like
phenomena is an effect of the assumed distributed reality rather
than its cause, any talk about interactions between primary
point-like phenomena is misplaced.  It is more meaningful to speak
of distributed interactions between inherently extended elements
of reality as having kinematic consequences for the corresponding
image locations.

\subsection{Examples of Nonlocality in Local Realist Wave Theories}
\label{ss:examples}

\subsubsection{Apparent Superluminal Light Propagation}

Culminating a series of similar results, a widely publicised
experiment by Wang et al \cite{Wang} featured a light pulse which
"transits" a chamber containing excited Caesium gas at an
effective speed of some 300 c.  The author's analysis, using local
realist wave theory, clearly illustrates the distinction between
ELOC and OLOC.

A pulse enters the chamber, and a pulse leaves the chamber, but
the pulse leaving is not the "same", in the sense of having a
persistent identity, as the one that entered. Instead, the result
is explained by phase disturbances amongst the wave components of
the input pulse induced by the dispersive medium, leading to a
reconstructed pulse at the output which appears ahead of the light
time through the apparatus. There is no contention that any
ontological element actually moved faster than light. Rather,
re-phasing of already distributed wave components causes the
observed non-locality. Although the physical scale of EPR
experiments is larger, the energies involved are relatively
trivial, so this mechanism can apply equally well to EPR, as
discussed in subsection \ref{ss:scenario}.

One important implication to note is that the distributed (wave
components) seem to have a better claim to the description "real"
than does the localised (image of the pulse).

\subsubsection{Tunnelling}

A variety of theoretical and experimental investigations have
shown the independence of tunnelling time with respect to the
dimensions of the barrier.  A more dramatic example however, is
provided in the case of a single particle tunnelling through two
successive barriers \cite{Olkohvsky}, wherein it is shown that the
traversal time depends neither upon the barrier widths, nor upon
the distance separating them.  Moreover, this paper emphasises the
formal analogy between the  Helmholtz and Schroedinger equations,
which enables the substitution of waveguide experiments (and the
associated local realist wave theoretical analysis) for particle
experiments. Again, re-phasing of already extended components
explains the non-locality, and there is no persistence of
identity.

\subsection{A Wave Theoretical Account of EPR}
\label{ss:scenario}

For context, let us consider an EPRB experiment involving two
electrons entangled in the angular momentum.  The entire
experiment comprises: the two particles, A and B; measurement
apparatus in each arm of the experiment, $M_{a}$ and $M_{b}$; and
the environment, W.  Let each of these subsystems be represented
by a suitable set of distributed wave components, designated as
$\{A\}$, $\{B\}$ and so on, all of which are assumed to extend in
space significantly farther than the distance separating the two
arms of the apparatus.

Now, under normal (unentangled) conditions, we find that
interactions are range dependent, so let us assume that,
immediately prior to any measurement having been performed, the
following sets of possible wave-wave interactions are
insignificant:

\noindent $\{A\}\leftrightarrow\{M_{b}\}\sim 0 \\
\{B\}\leftrightarrow\{M_{a}\}\sim 0 \\
\{M_{a}\}\leftrightarrow\{M_{b}\}\sim 0$

Where the symbol $\leftrightarrow$ means "exchanges of  momentum
between".

Ordinarily we would also expect $\{A\}\leftrightarrow \{B\}\sim
0$, however, the particles having been created at a common origin
with zero nett angular momentum, their respective sets of wave
components are intimately correlated which provides a legitimate
physical basis upon which to anticipate a continuing interaction
sufficient, under carefully controlled experimental conditions, to
maintain the entanglement in spite of some modest, decohering
interactions with the environment, $\{W\}\leftrightarrow\{A\}$ and
$\{W\}\leftrightarrow\{B\}$.

The close range measurement interaction,
$\{A\}\leftrightarrow\{M_{a}\}$ or
$\{B\}\leftrightarrow\{M_{b}\}$, whichever happens first (in the
MBR frame), predominates over the $\{A\}\leftrightarrow \{B\}$
interaction, disrupting the interparticle correlations and
therefore terminating any future possiblity of a significant
interaction between the two particles.  Note that the wave
components $\{M_{a}\}$ are widely distributed, as are the wave
components $\{A\}$, so there is no paradox in the proposition that
this disruption happens instantaneously at large distances from
the location of the particle / measurement equipment, as recently
pointed out:

\begin{quote}

"Of course, if we make some excitation for field at the point O,
then a propagation of this excitation from this point will have a
finite speed.  But in the scope of the unified field model we do
not be able to make this excitation or modify arbitrarily the
world solution.  Any excitations of the field at the point O
belong to the world solution which is a single whole" (sic) - A.A.
Chernitski \cite{Chernitski}

\end{quote}

In a wave theory local realism means that changes propagate away
from their causes at (or below) the speed of light, but there are
no point sources, only distributed excitations.

Prior to any measurement interaction, then, the two particles
co-evolve. Each one might\footnote{or might not, if we think about
the $\{A\}\leftrightarrow\{B\}$ interaction as a slight resonance,
storing energy neither in one particle nor the other, but in the
combined distributed system. In either case, the realist sense is
maintained.} be said to have at any instant a certain angular
momentum, in balance with the other member of the pair, and the
projection of this onto the selected measurement axis determines
the distribution of outcomes for the measured
particle\footnote{Applicable (non-local) Hidden Variables models
exist, however, as repeatedly stressed, there is no motivation to
depart from the restriction to observables - our arguments are
only intended to show that local realism is conceptually
compatible with Quantum Mechanical nonlocality.}. Angular momentum
conservation then requires that the unmeasured particle is left in
the corresponding eigenstate.

There is no "spooky action at a distance" involved, and the
essential feature of the experiment is the distributed cessation
of a distributed interaction that had been ongoing prior to the
first measurement, all of which provides a physical sense to the
quantum mechanical concept of wave function collapse.

\section{Conclusions}\label{sec:conc}

Recent EPR experiments provide increasingly secure experimental
evidence that there are indeed influences between space like
separated events. Further evidence to the same effect has been
found in the context of gravity \cite{TVF}, where astronomical
observations establish a minimum bound of $\sim{10^{10}c}$ for the
velocity of any putative gravitational IMP.

Now, since it is customary to seek to replace a successful
physical theory when, and only when, it is found to be empirically
deficient, the fundamental challenge presented by EPR is not to
improve upon quantum mechanics, but to identify and develop a
world view commensurate with it. There are two principal
impediments to such a development.

The first of these, introduced no less than two and a half
millenia ago, is the atomist metaphysic, which has it that the
world ultimately consists of a multitude of distinct elementary
particles. Whilst this approach is appealing to the extent that it
mimics our ordinary experience of a world apparently divided into
spatially distinct objects, it fails to provide an adequate
account of interaction between remote particles. Retarded
interaction has long been thought to overcome the indisputable
logical necessity that interaction must be predicated upon
co-location. However, this metaphysical bandaid addresses neither
the question of how the source particle "knows" which Virtual IMPs
have actually been absorbed (in order to account for the momentum
transferred to the target), nor the fact that the co-location
constraint is never satisfied by a point model (because, if we
treat them both as point particles, no matter how close the
mediating particle may be to the target, it remains separated from
it by an infinite number of points).  Therefore, the elements of
any ontological proposition competent to represent interaction
processes must be inherently extensive. Concerning the degree of
extension, there are two possibilities: ontological elements may
be effectively bounded (e.g. within a radius of
$\sim{10^{-15}\textrm{m}}$ for an electron) or effectively
unbounded, as in this article.

In the bounded case, retardation is still a requirement, so EPR
mandates that we need to invoke not just IMPs, but superluminal
ones at that.  Such an approach seems quite unrelated to the
foundation theories. Why should mechanics be non-linear? Why
should massive particles projected onto a slit system display
interference patterns?  How can the source particle know when, or
if, a given virtual IMP is absorbed? The conceptual gap between
the foundations remains unbridged. Finally, the pseudo-point
particle approach to ontology is profligate: we have propagating
and non-propagating, inherently subluminal, inherently photonic
and inherently superluminal classes of phenomena, with no means to
comprehend transitions amongst these classes.

Contrasting with this, we have considered the unbounded case with
ontological elements constrained to propagate at the
characteristic velocity. The restriction to extensive elements
immediately poses the question of how it is that phenomena appear
to be well-localised, which points directly to Fourier
superposition methods in general, and wave solitons in particular.
There is only one ontological class, so there is no question of
unexplained inter-class transitions. Under the assumed dynamical
constraint, a wave theoretical approach to particle mechanics
reproduces the Special Theory and all its valuable symmetries,
with the one essential exception that the theory is automatically
embedded in the only suitable candidate preferred frame that has
been identified experimentally - a critical fact in the light of
EPR's instantaneous correlations, but one that cannot be explained
within the usual interpretation of Relativity Theory.  Taking our
locally realistic wave elements to be widely distributed allows
the model to embrace apparently instantaneous interaction at a
distance, just as has been found with the classical local realist
wave theory.  Qualitatively, this approach is clearly compatible
with quantum mechanics.  Quantitatively, the invariance of the
3-velocity in a physical wave model implies the invariance of the
4-velocity for particles, which in turn has been shown to imply
most of the physics presently in use, including quantum mechanics'
most important founding equations.

Is this not the simplest resolution of the EPR paradox imaginable?

The second impediment to the development of a world view
commensurate with the state of the art in Physics lies somewhere
in the no man's land between Physics and Metaphysics.  The single
most vital practical principle underlying the successful
development of Physics to date has been the restriction to
observables.  This is what allows experiments to be the ultimate
arbiters of our models.  However, as helpful as it is for Physics,
it can be equally unhelpful to Metaphysics, because we can never,
in-principle, observe reality sui generis, only properties. The
very process of observation is necessarily a matter of abstracting
from the reality.  Therefore one should not proceed too
confidently from Physics to Metaphysics, from the truth value of
an empirical model to truths about the ultimate nature of physical
reality.

Unfortunately, it must be said that this is precisely what has
occurred with Relativity Theory: an empirically valid fact (that
observers will not necessarily agree about the temporal ordering
of spatially separated events), has combined with the presumption
of atomism to evolve into an incontrovertible metatruth (that
there is no fact of the matter concerning such temporal
orderings). We have moved, with no positive proof, from an
empirical model with the attraction that it does not rely upon the
preferred frame to the certain metaphysical position that there
cannot be one.

Can relativity of simultaneity, having been discovered by pure
reason rather than by any mundane experiment, be so absolute as to
transcend multiple forms of empirical evidence (EPR, gravity, and
an experimental preferred frame)? Should Physics disavow all
metaphysical pretensions yet take its own deductions to be
metaphysically axiomatic? And if we must continue to cling to a
philosophical position against the evidence, mounting for more
than thirty years now, what became of the restriction to
observables?

\newpage

\begin{figure}[h]
\begin{picture}(80,280) (-55,0)
\includegraphics[width=120 mm, height=100mm] {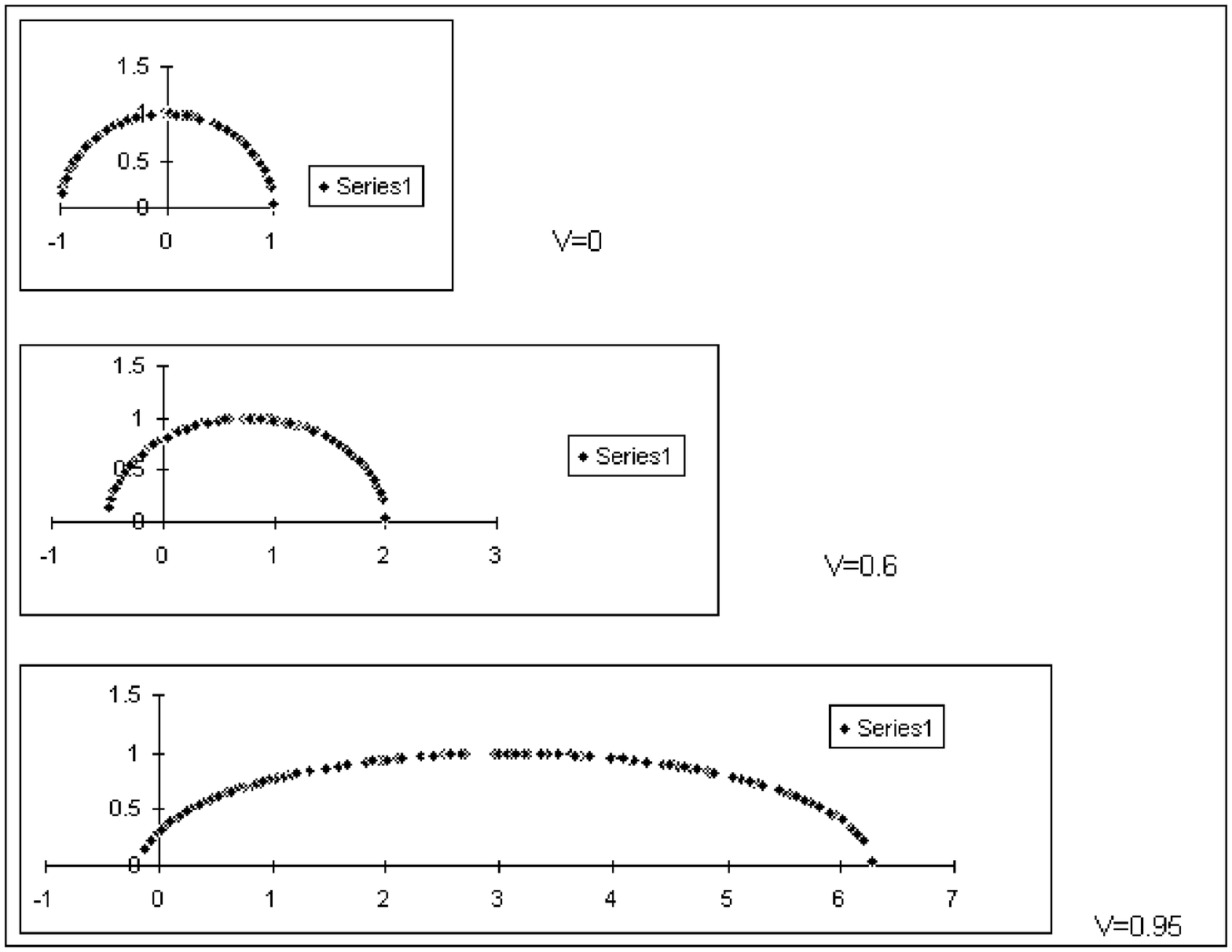}
\end{picture}
\caption{Momentum Flux Distributions at various particle
velocities. Each plot is an ellipsoid of revolution about the
X-axis, which lies in the direction of motion.  Eccentricity
increases with the group velocity, and the geometric centre of any
given ellipsoid corresponds to the particle momentum.}
 \label{fig1}
\vspace{2cm}
\begin{picture}(100,150) (-70,+00)
\includegraphics[width=100 mm, height=50mm] {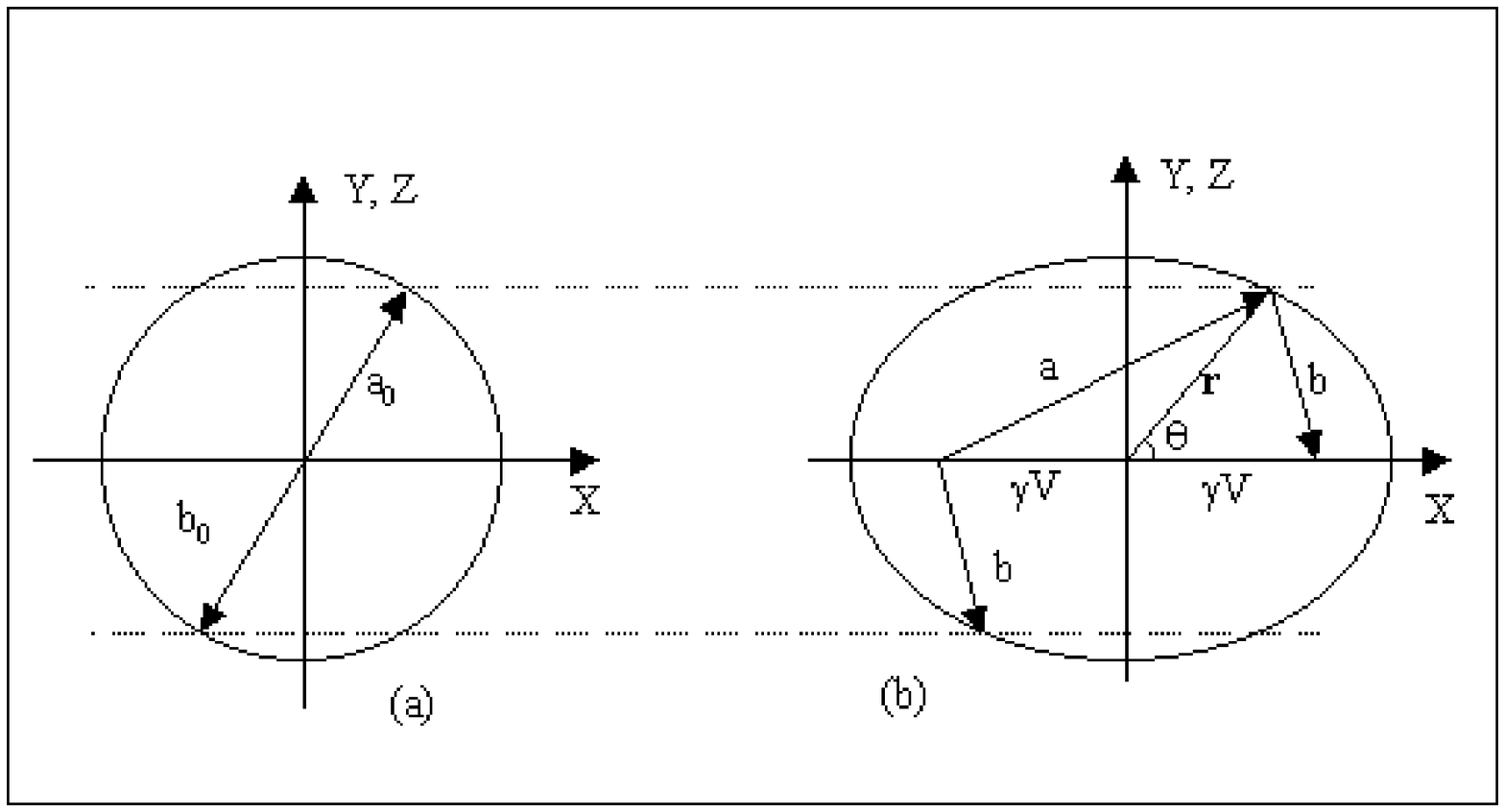}
\end{picture}
\caption{Transformation of the momentum flux distribution.  Waves
$\textrm{a}_{0}, \textrm{b}_{0}$ map to waves a, b respectively a)
Rest particle; b) group velocity = $
\textrm{V}_{x}\widehat{\textbf{i}}  $}
 \label{fig2}
\end{figure}


\begin{thebibliography}{widest-label}

\bibitem{EPR}    Einstein A., Podolsky B. and Rosen N. \ \ Phys. Rev.
47, 777 (1935)

\bibitem{Bell}  Bell J.S.  \ \ Physics 1, 195 (1965)

\bibitem{Clauser} Clauser J. and Horne M. \ \ Detector Inefficiencies
in the EPR experiment. \ \  Phys. Rev. D 35 (12) 3831-3835 (1987)

\bibitem{Vaidman} Vaidman \ \ L. Tests of Bell Inequalities \ \
quant-ph/0107057   (2001)

\bibitem{Aspect}  Aspect A., Grangier P. and Roger B. \ \ Phys. Rev.
Lett. 49, 1804-1807 (1982)

\bibitem{Scarani}  Scarani V., Tittel W., Zbinden H., Gisin N. \ \
quant-ph/0007008 \ \  Phys. Lett. A 276 2000 1-7 \ \ (2000)

\bibitem{Ou}   Ou Z.Y., Pereira S.F., Kimble H.J. and Peng K.C. \ \
Phys Rev. Lett. 68, 3663 (1992)

\bibitem{Percival} Percival I.C. \ \  Physical Letters A  244: (6) p495-501 (1998)

\bibitem{Smoot}    Smoot G. \ \   Detection of Anisotropy in the Cosmic
Blackbody Radiation. \ \ Physical Review Letters  39  14  p898
(1977)

\bibitem{Peebles3}    Peebles P.J.E. and Wilkinson D.T. \ \ Phys. Rev. 174 2168
(1968)

\bibitem{Longair}    Longair M.S. \ \   The Physics of Background Radiation.  The
Deep Universe.  Springer Verlag (1995)

\bibitem{Weinberg}    Weinberg S. \ \ The First Three Minutes.  Basic Books.
p71-72.  (1977)

\bibitem{Dirac} Dirac P.A.M.  \ \  The Theory of the electron (parts
1 and 2). \ \ Proceedings of the Royal Society in London.  A117,
p610 and A118,  p351 (1928)

\bibitem{Breit} Breit G.  \ \  An interpretation of Dirac's Theory of the
electron. \ \  Proceedings of the National Academy of Sciences USA
14 p553 (1928)


\bibitem{schroedinger} Schroedinger E. \ \ Sitzungsberichte Berlin Akadamie \ \ p418 (1930)

\bibitem{Dirac3} Dirac P.A.M.  \ \  Principles of Quantum Mechanics.  Oxford. p
263. (1958)

\bibitem{Messiah}    Messiah A.   \ \   Quantum Mechanics.  Vol 2. Ch XX pp
922-925

\bibitem{Bohm} Bohm D.  \ \  The Special Theory of Relativity. pp
23-25.  (1996)

\bibitem{Hafele} Hafele J. and Keating R. \ \ Science 177 p166  (1972)

\bibitem{Kundig}    Kundig W. \ \  Phys Rev. 129, 2371 (1963)

\bibitem{Ives}    Ives H. and Stillwell. \ \ Journal of the Optical Society of
America 28 \ \ pp 215-226 (1938) and 31 p369 (1941)

\bibitem{Konopinski2}    Konopinski E.J. \ \  Electromagnetic fields and Relativistic
particles \ \  McGraw Hill  p315-319  (1981)

\bibitem{Poynting}    Poynting J.H. \ \  Phil. Trans. 175 p343-361  (1884)

\bibitem{BohmHiley} Bohm D. and Hiley B. \ \ The Undivided Universe:
an ontological interpretation of quantum mechanics. Routledge. \ \
pp288-292  (1993)


\bibitem{Teller}    Teller P.  \ \   An Interpretive Introduction to Quantum
Field Theory.  Ch 7 (1995)

\bibitem{Cui2}  Cui H.Y. \ \ Direction Adaptation Nature of Coulomb's Force and
Gravitational Force in 4-Dimensional Space-time \ \
physics/0102073 (2001)

\bibitem{Cui1}  Cui H.Y. \ \ Method for Deriving the Dirac Equation from the Relativistic
Newton's Second Law \ \  quant-ph/0102114 (2001)

\bibitem{Redhead}  Redhead M. \ \ Incompleteness, Nonlocality and
Realism. \ \ Oxford University Press  (1987)

\bibitem{SL1}  Rembielinski J. \ \ Superluminal Phenomena and the Quantum Preferred
Frame \ \ quant-ph/0010026 (2000)

\bibitem{Konopinski} Konopinski E. J. \ \ Electromagnetic Fields and Relativistic
Particles.  Mcgraw Hill \ \ pp 441-454  (1981)

\bibitem{d'Espagnat}  d'Espagnat B. \ \ quant-ph/0101141 (2001)

\bibitem{Wang}  Wang L.J., Kuzmich A., Dogariu A. \ \ Nature 406 pp 277-279 (2000)

\bibitem{Olkohvsky}  Olkhovsky V. S., Recami E. and Salesi G. \ \  Superluminal
effects for quantum tunnelling through TWO successive barriers \ \
quant-ph 0002022 v4 (2001)

\bibitem{Chernitski} Chernitskii A.A. \ \ Concept of Unified Local Field Theory and
Nonlocality of Matter \ \ quant-ph/0102101 (2001)

\bibitem{TVF}  Van Flandern T. \ \ The Speed of Gravity - What the Experiments
Say.\ \ Physics Letters A, 250 1-11  (1998).



\end{thebibliography}
\end{document}